\documentclass[twocolumn,aps,rmp,floats]{revtex4}
\usepackage{graphics,graphicx,epsfig}
\usepackage{amssymb}
\usepackage{epsf,epstopdf,wrapfig}

\begin{document}

\title{Weak pairwise correlations imply strongly correlated\\ network states in a neural population}

\author{Elad Schneidman,$^{a-c}$  Michael J. Berry II,$^b$ Ronen Segev$^b$ and William Bialek$^{a,c}$}

\affiliation{$^a$Joseph Henry Laboratories of Physics,
$^b$Department of Molecular Biology, and
$^c$Lewis--Sigler Institute for Integrative Genomics,
Princeton University, Princeton, New Jersey 08544 USA}

\date{\today}

\begin{abstract}
Biological networks have so many possible states that exhaustive sampling  is impossible. Successful analysis thus depends on simplifying hypotheses, but experiments on many systems hint that complicated, higher order interactions among large groups of elements play an important role. In the vertebrate retina,  we show that weak correlations between pairs of neurons coexist with strongly collective behavior in the responses of ten or more neurons.  Surprisingly, we find that this collective behavior  is described quantitatively by models   that capture the observed pairwise correlations but assume no higher order interactions.  These maximum entropy models are equivalent to Ising models, and predict that larger networks are completely dominated by correlation effects.  This suggests  that the neural code  has  associative or error- correcting properties, and we provide preliminary evidence for such behavior. As a first test for the generality of these ideas, we show that similar results are obtained from networks of cultured cortical neurons.
\end{abstract}

\maketitle

Much of what we know about biological networks has been learned by studying one element at a time---recording the electrical activity of single neurons,  the expression levels of single genes or the concentrations of individual metabolites.  On the other hand, important aspects of biological function must be shared among many elements \cite{hopfield+tank_86,georgopolous+al_86,hartwell+al_99,barabasi+oltvai_04}.   
As a first step beyond the analysis of elements in isolation, much attention has been focused on the pairwise correlation properties of these elements, both in networks of neurons \cite{perkel+bullock_68,Zohary+al_94,meister+al_95, riehle+al_97,dan+al_98,hatsopolous+al_98,Abbott+Dayan_99, bair+al_01,Shamir+Sompolinsky_04} and in networks of genes \cite{eisen_98,alter+al_00,holter+al_01}.  
But given a characterization of pairwise correlations, what can we really say about the whole network?  How can we tell if  inferences from a pairwise analysis are correct, or if they are defeated by  higher order interactions among triplets, quadruplets, ... of elements?  If these effects are important, how can we escape from the `curse of dimensionality' that arises because there are exponentially many possibilities for such terms?

Here we address these questions in the context of the vertebrate retina, where it is possible to make long, stable recordings from many neurons simultaneously as the system responds to complex, naturalistic inputs \cite{markus1,EJ,data_a,data_b}. We compare the correlation properties of cell pairs with the collective behavior in larger groups of cells, and find that the minimal model which incorporates the pairwise correlations provides strikingly accurate but non--trivial predictions of the collective effects.  These minimal models are equivalent to the Ising model in statistical physics, and this mapping allows us to explore the properties of larger networks, in particular their capacity for error--correcting representations of incoming sensory data.\footnote{A preliminary account of this work was presented at the conference on Computational and Systems Neuroscience (COSYNE), 17--20 March 2005, in Salt Lake City, Utah.  See {\tt http://cosyne.org}.}

\section{The scale of correlations}

Throughout the nervous system, individual elements communicate by generating discrete pulses termed action potentials or spikes \cite{spikes}.  If we look  through a window of fixed time resolution $\Delta \tau$, then for small $\Delta \tau$ these responses are binary---either the cell spikes (`$1$') or it doesn't (`$0$').
Although some pairs of cells  have very strong correlations, most correlations are weak, so that the probability of seeing synchronous spikes is almost equal to the product of the probabilities of seeing the individual spikes; nonetheless, these weak correlations are statistically significant for most if not all pairs of nearby cells.  All of these features are illustrated quantitatively by an experiment on the salamander retina (Fig. 1), where we record simultaneously from 40 retinal ganglion cells 
as they respond to movies taken from a natural setting (see Methods).
Mean spike rates range from 0.3 to 4.5 spikes/s and the correlations between cells have structure on the scale of $\Delta \tau = 20\,{\rm ms}$.  Using this window to define binary variables, we find correlation coefficients that range from $-0.03$ to $0.5$, with ninety percent of the results falling in the narrow range $C \in [-0.02, +0.1]$.

\begin{figure*}
\vskip -1.75 in
\centerline{\psfig{figure=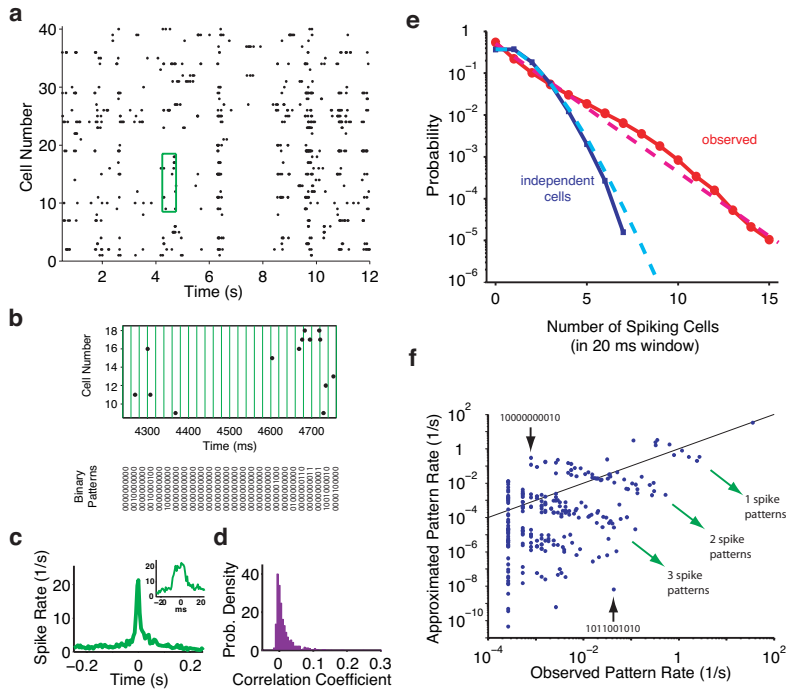,width=1.8\columnwidth}}
\vskip  -2.25 in
\caption{Weak pairwise cross--correlations and the failure of the independent approximation.
(a) A segment of the simultaneous responses of 40 retinal ganglion cells in the salamander to a natural movie clip. Each dot represents the time of an action potential. (b) Discretization of population spike trains into a binary pattern is shown for the green boxed area in panel a: spike trains are binned using $\Delta \tau=20\,{\rm ms}$ windows (top panel), into binary sequences of spiking (1) and non-spiking (0); in the rare cases where there is more than one spike in a bin, we denote it as `1'. Every string (bottom panel) describes the activity pattern of the cells on a give time point.  For clarity, 10 out of 40 cells are shown.
(c) Example cross--correlogram between two neurons with strong correlations;  the average firing rate of one cell is plotted relative to time at which the other cell spikes. Inset shows the same cross--correlogram on an expanded time scale.
(d) Histogram of correlation coefficients for all pairs of 40 cells from panel a.  We discretize the neural response into binary (spike/no spike) variables for each cell, using $\Delta\tau = 20\,{\rm ms}$ bins as in (b), and then compute the correlation coefficients among these variables. Because the data sets we consider here are very large ($\sim 1\,$hour), the threshold for statistical significance of the individual correlation coefficients is well below $|C| = 0.01$.
(e) Probability distribution of synchronous spiking events in the 40 cell population in response to a long natural movie (red) approximates an exponential  (dashed red).  The distribution of synchronous events for the same 40 cells after shuffling each cell's spike train to eliminate all correlations (blue), compared to the  Poisson distribution  (dashed light blue).  (f) The rate of occurrence of each pattern predicted  if all cells are independent is plotted against the measured rate.  Each dot stands for one of the $2^{10} = 1024$ possible binary activity patterns for 10 cells. Black line shows equality. Two examples of extreme mis--estimation of the actual pattern rate by the independent model are highlighted (see text). The logarithmic scale emphasizes the large dynamic range and large errors of the independent approximation; less clear is that the (very accurately measured) probability of the response $0000000000$ is underestimated by $\sim 7\%$.}
\end{figure*}

The small values of the correlation coefficients  suggest an approximation in which the cells are completely independent.  For most pairs, this  is true with a precision of a few percent, but if we extrapolate this approximation to the whole population of 40 cells, it fails disastrously.  In Fig. 1e we show the probability $P(K)$ that $K$ of these cells generate a spike in the same small window of duration $\Delta \tau$.  If the cells were independent, $P(K)$ would approximate the Poisson distribution, while the true distribution is  nearly exponential.  The probability of $K=7$ cells spiking together is $\sim 10^3\times$ larger than expected in the independent model, and at $K=10$ the discrepancy is $\sim 10^5$:  10--spike events occur several times per minute in this small patch of the retina, while  the independent model predicts that they should occur once every three weeks.  

The discrepancy between the independent model and the actual data is even more clear if we look at particular patterns of response across the population.  Choosing $N=10$ cells out the 40, we can form an $N$--letter binary word to describe the instantaneous state of the network, as in Fig 1b.  
The independent model makes simple predictions for the rate at which each such word should occur, and Fig.1f shows these predictions as a scatter plot against the actual rate at which the words occur in the experiment. At one extreme, the word $1011001010$ occurs once per minute, while the independent model predicts that this should occur once per three years.  Conversely,  the word $1000000010$ is predicted to occur once per three seconds, while in fact it occurred only three times in the course of an hour.  Although these are extreme cases,  the independent model makes order--of--magnitude errors   even for the rates of very common patterns of activity, such as a single cell generating a spike while all others are silent.  Indeed, in the scatter plot of predicted vs. observed rates we see clusters corresponding to different total numbers of spikes, but within each cluster the predictions and observations are strongly anti--correlated.

We conclude that  weak correlations among pairs of neurons coexist with strong correlations in the states of the population as a whole.   
One possible explanation is that there are specific multi--neuron correlations, whether driven by the stimulus or intrinsic to the network,  which simply are not measured by looking at pairs of cells. Searching for such higher order effects presents many challenges  \cite{Martignon+al-00,Grun+al_02,Schnitzer-Meister-03}. 
Another scenario is that small correlations among very many pairs could add up to  a strong effect on the network as a whole.  If correct, this would be an enormous simplification in our description of the network dynamics.

\section{Minimal consequences of pairwise correlations}

To describe the network as a whole we need to write down a probability distribution for the $2^N$ binary words corresponding to patterns of spiking and silence in the population.    
The pairwise correlations tell us something about this distribution, but there are an infinite number of models that are consistent with a given set of pairwise correlations. The difficulty thus is to find a distribution that is consistent {\em only} with the measured correlations, and does not implicitly  assume the existence of unmeasured higher--order interactions.     Since the entropy of a distribution measures the randomness or lack of interaction among different variables \cite{brillouin}, this minimally structured distribution that we are looking for is the maximum entropy distribution \cite{Jaynes-57} consistent with the measured properties of individual cells and cell pairs \cite{schneidman+al_03}.

We recall that maximum entropy models have a close connection to statistical mechanics:  physical systems in thermal equilibrium are described by  the Boltzmann distribution, which has the maximum possible entropy given the mean energy of the system \cite{Jaynes-57,LL}.  Thus any maximum entropy probability distribution defines an energy function  for the system we are studying, and we will see that  the energy function relevant for our problem is an Ising model.  Ising models have been discussed  extensively as models for neural networks \cite{hopfield_82,amit_89}, but in these  discussions the model arose from specific hypotheses about the network dynamics.  Here the Ising model is forced upon us as the least structured model that is consistent with measured  spike rates and pairwise correlations; we emphasize that this is not an analogy or a metaphor, but rather an exact mapping.

Whether we view the maximum entropy model through its analogy with statistical physics or simply as a model to be constructed numerically from the data (see Methods), we need meaningful ways of assessing whether this model is correct.  Generally, for a network of $N$ elements, we can define maximum entropy distributions $P_K$ that are consistent with all $K^{\rm th}$--order correlations for any $K=1, 2, \cdots , N$  \cite{schneidman+al_03}.  
These distributions  form a hierarchy,   from $K=1$ where all elements are independent,  up to $K=N$, which is an exact description that allows arbitrarily complex interactions; their entropies $S_K$ decrease monotonically toward the true entropy $S$:  $S_1 \geq S_2 \geq \cdots \geq S_N = S$.  
The entropy  difference or multi--information $I_N = S_1 - S_N$ measures the total amount of correlation in the network, independent of whether it arises from pairwise, triplet, or more complex correlations \cite{cover}. The contribution of the $K^{\rm th}$--order correlation is  $I_K = S_{K-1} - S_{K}$ and is always positive (more correlation always reduces the entropy);  $I_N$ is the sum of all the $I_K$ \cite{schneidman+al_03}.
Therefore, the question of whether pairwise correlations provide an effective
description of the system becomes the question of whether the reduction in entropy that comes from these correlations, 
$I_2 = S_1 - S_2$, captures all or most of the multi--information $I_N$.

\section{Are pairwise correlations enough?}

\begin{figure*}
\vskip -1.65 in
\centerline{\psfig{figure=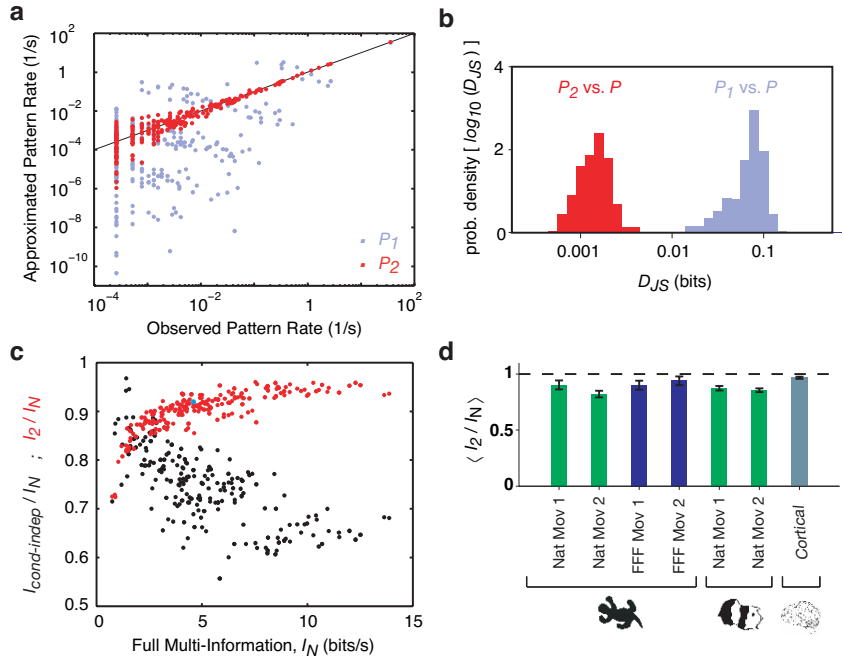,width=1.8\columnwidth}}
\vskip  -2.5 in
\caption{A maximum entropy model including all pairwise interactions gives an excellent approximation of the full network correlation structure.
(a) Using the same group of 10 cells from Fig. 1, the rate of occurrence of each firing pattern predicted from the the maximum entropy model $P_2$ that takes into account all pairwise correlations is plotted against the measured rate (red dots).
The rates of commonly occurring patterns are predicted with better than $10\%$ accuracy, and scatter between predictions and observations is confined largely to rare events for which the measurement of rates is itself uncertain.
For comparison, the independent model $P_1$ is also plotted (from Fig. 1f; gray points). Black line shows equality.
(b) Histogram of Jensen--Shannon divergences between the actual probability distribution of activity patterns in 10 cell groups and the models $P_1$ (gray) and $P_2$ (red); data from 250 groups.
Note that the histograms are plotted on a logarithmic scale, because the two cases have values differing by two orders of magnitude.  
(c) Fraction of full network correlation in 10 cell groups that is captured by the maximum entropy model of 2nd order, $I_2 / I_N$, plotted as a function of the full network correlation, measured by the multi--information $I_N$ (red dots). The multi--information values are multiplied by $1/\Delta \tau$ to give bin--independent units. Every dot stands for one group of 10 cells.  The 10 cell group featured in panel a is shown as an light blue dot. For the same sets of 10 cells, the fraction of information  of full network correlation in 10 cell groups that is captured by the conditional independence model, $I_{\rm cond-indep} / I_N$, is shown in black (see text).
(d) Average values of $I_2/I_N$  from 250 different 10 cell groups.
Results are shown  for different movies, for different species, and for cultured cortical networks; error bars show standard errors of the mean.  Similar results are obtained on changing $N$ and $\Delta\tau$.} 
\end{figure*}

Figure 2 shows the predictions of the maximum entropy model $P_2$ consistent with pairwise correlations in populations of $N=10$ cells.
Looking in detail at the patterns of spiking and silence of one group of 10 cells, we see that the predicted rates for different binary words are tightly correlated with the observed rates over a very large dynamic range, so that the dramatic failures of the independent model have been overcome (Fig 2a).

With 40 cells we can choose many different populations of 10 cells, and in each case we find that the predicted and observed distributions of words are very similar.  This is quantified by the Jensen--Shannon divergence $D_{JS}$ \cite{lin_91}, which essentially measures the inverse of the number of independent samples we would need in order to be sure that the two distributions were different. 
While the true distribution of responses  is not exactly the same as that predicted by the maximum entropy model, it typically would take thousands of independent samples to distinguish reliably between them, two orders magnitude more than in the independent model  (Fig. 2b).

The success of the pairwise maximum entropy models in capturing the correlation structure of the network is summarized by the fraction $I_2/I_N \sim 90\%$ (Fig 2c). This ratio is larger when $I_N$ itself is larger, so that the pairwise model is more effective in describing populations of cells with stronger correlations, and
the ability of this model to capture $\sim 90\%$ of the multi--information holds independent of many details (Fig 2d): 
We can vary the particular natural movies shown to the retina, use an artificial movie, change the size of the bins $\Delta\tau$ used to define the binary responses,  the number of neurons $N$ that we analyze, and even shift from a lower vertebrate (salamander) to a mammalian (guinea pig) retina.   Finally,  the correlation structure in a network of cultured cortical neurons \cite{marom} can be captured by the pairwise  model with similar accuracy.

The maximum entropy model describes the correlation structure of the network activity without assumptions about its mechanistic origin.  A more traditional approach has been to dissect the correlations into contributions that are intrinsic to the network and those that are driven by the visual stimulus. The simplest model in this view is one in which cells spike independently in response to their input, so that all correlations are generated by covariations of the individual cells' firing rates \cite{Perkel+al_67}.  Although there may be situations in which conditional independence is a good approximation, Fig 2c shows that this model is less effective than the maximum entropy model in capturing the multi--information for 232/250 groups of 10 neurons.  The hypothesis of conditional independence is consistently less effective in capturing the structure of more strongly correlated groups of cells, which is opposite to the behavior of the maximum entropy model.
Note also that because each cell has its own spike rate, potentially different at each moment in time, the conditionally independent model has has $NT/\Delta \tau$ parameters, where $T$ is the duration of the stimulus movie; in our case $NT/\Delta \tau\sim 10^4$, in contrast to the $N(N+1)/2 = 55$ parameters of the maximum entropy model.   Finally,  while the maximum entropy model can be constructed solely from the observed correlations among neurons, the conditionally independent model requires explicit access to repeated presentations of the visual stimulus.  
Thus while the central nervous system could learn the maximum entropy model from the data provided by the retina alone, the conditionally independent model is not biologically realistic in this sense.

We conclude that although the pairwise correlations are small and the multi--neuron deviations from independence are large, the maximum entropy model consistent with the pairwise correlations captures almost all of the structure in the distribution of responses from the full population of neurons.  Thus,  the weak pairwise correlations {\em imply}  strongly correlated states. To understand how this happens, it is useful to look at the mathematical structure of the maximum entropy distribution.

 \section{Ising models, revisited}

\begin{figure}
\vskip -0.5 in
\centerline{\psfig{figure=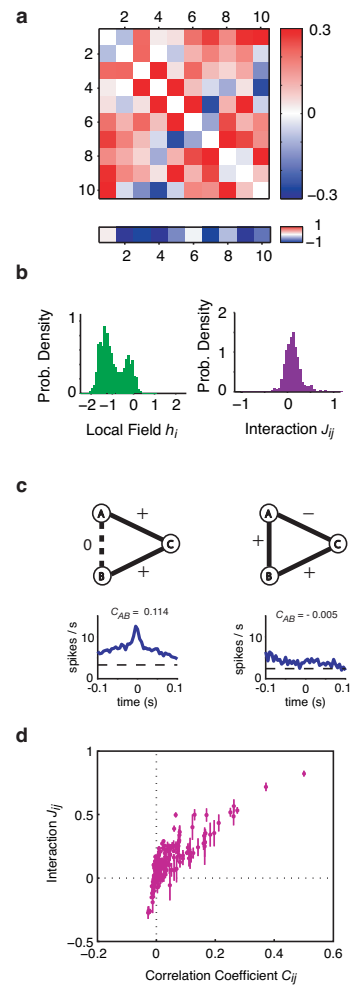,width=1.5\columnwidth}}
\vskip 0.1 in
\caption{Pairwise interactions and individual cell biases, as in Eq. (\ref{ising1}).
(a) Example of the pairwise interactions $J_{\rm ij}$ (above) and bias values (or local fields) $h_{\rm i}$ (below) for one group of 10 cells.  (b) Histograms of $h_{\rm i}$ and $J_{\rm ij}$ values  from 250 different groups of 10 cells.
(c) Two examples of 3 cells within a group of 10.  At left, cells   $A$ and $B$ have almost no interaction ($J_{\rm AB}=-0.02$), but cell $C$ is very strongly interacting with both $A$ and $B$ ($J_{\rm AC}=0.52$,$J_{\rm BC}=0.70$), so that cells $A$ and $B$ exhibit strong correlation, as shown by their cross--correlogram (bottom panel; $C_{\rm AB}=0.11$). At right, a  ``frustrated'' triplet,  in which cells $A$ and $B$ have a significant positive interaction ($J_{\rm AB}=0.13$), as do cells $B$ and $C$ ($J_{\rm BC}=0.09$), but $A$ and $C$ have a significant negative interaction ($J_{\rm AC}=-0.11$).  As a result, there is no clear correlation between cells $A$ and $B$, as shown by their cross--correlogram (bottom panel; $C_{\rm AB}=-0.005$). (d) Interaction strength $J_{\rm ij}$ plotted against the correlation coefficient $C_{\rm ij}$; each point shows the value for one cell pair averaged over many different groups of neighboring cells (190  pairs from 250 groups), and  error bars show standard deviations.
}
\end{figure}

We recall that the maximum entropy distribution consistent with a known average energy $\langle E\rangle$ is the Boltzmann distribution, $P\propto \exp(-E/k_B T)$.  This generalizes so that if we know the average values of many variables $f_\mu$describing the system, then the maximum entropy distribution is $P \propto \exp(-\sum_\mu\lambda_\mu f_\mu)$, where there is a separate Lagrange multiplier $\lambda_\mu$ for each constraint \cite{Jaynes-57,LL}.  
In our case, we are given  the average probability of a spike in each cell and the correlations among all pairs.  If we represent the activity of cell $\rm i$ by a  variable $\sigma_{\rm i} =\pm1$, where $+1$ stands for spiking and $-1$ stands for silence, then these constraints are equivalent to fixing  the average of each $\sigma_{\rm i}$ and the averages of all products $\sigma_{\rm i}\sigma_{\rm j}$, respectively.  The resulting maximum entropy distribution is
\begin{eqnarray}
P_2(\sigma_1\hskip-0.12in &,& \hskip -0.1in \sigma_2 ,  \cdots , \sigma_N ) \nonumber\\
&=& {1\over Z} \exp\left[ \sum_{\rm i} h_{\rm i}\sigma_{\rm i} +{1\over 2} \sum_{\rm i\neq j}J_{\rm i j} \sigma_{\rm i}\sigma_{\rm j} \right] ,
\label{ising1}
\end{eqnarray}
where the Lagrange multipliers $\{ h_{\rm i} , J_{\rm ij}\}$ have to be chosen so that the averages $\{\langle \sigma_{\rm i} \rangle , \langle \sigma_{\rm i}\sigma_{\rm j}\rangle\}$ in this distribution agree with experiment. 
This is the Ising model \cite{LL}, where the $\sigma_{\rm i}$ are spins, the $h_{\rm i}$ are local magnetic fields acting on each spin, and the $J_{\rm ij}$ are the exchange interactions;
note that $h >0$ favors spiking and $J > 0$ favors positive correlations.

Figure 3 shows  the parameters $\{ h_{\rm i} , J_{\rm ij}\}$ for a particular group of ten cells, as well as the distributions of parameters for many such groups. 
Most cells have a negative local field, which biases them toward silence.  Figures 3c and d illustrates the non--trivial  relationship between the pairwise interaction strengths $J_{\rm ij}$  and the observed pairwise correlations.  
Correlations can be induced between two cells in the absence of direct interaction when both cells interact with a common neighbor (Fig 3c, left).
Alternatively, if the interactions among pairs of cells have alternating signs, then 
it is possible to have 
``frustration''  among triplets of cells destroy the correlations between two cells even when they have a strong positive interaction (Fig 3c, right).
Roughly  $40\%$ of all triangles are frustrated in this way, and we return to this below.

We can rewrite Eq (\ref{ising1}) exactly by saying that each neuron or spin $\sigma_{\rm i}$ experiences an effective magnetic field that includes the local field or intrinsic bias $h_{\rm i}$ and a contribution from interactions with all the other spins, $h_{\rm i}^{\rm int} = {1\over 2}\sum_{\rm j\neq i} J_{\rm ij} \sigma_{\rm j}$; note that $h_{\rm i}^{\rm int}$ depends on whether the other cells are spiking or silent.
The intrinsic bias dominates in small groups of cells, but as we look to larger networks, the fact that almost all of the $\sim N^2$ pairs of cells are significantly (if weakly) interacting shifts the balance so that the typical values of the intrinsic bias are reduced while the effective field contributed by the other cells has increased (Fig. 4).  Intuitively,  although the influence of one cell on another is small, 
the summed effect of many cells on one cell can be large, so that states of the whole network are far from those predicted by the independent approximation.   

\begin{figure}
\begin{centering}
\vskip -1.75 in
\centerline{\psfig{figure=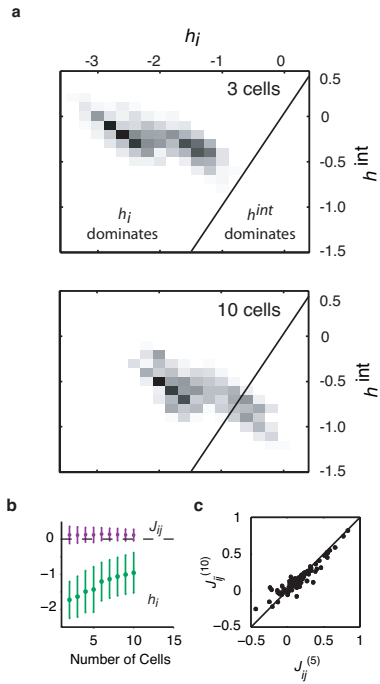,width=1.6\columnwidth}}
\vskip -1.3 in
\caption{Interactions and local fields in networks of different size. (a) Grayscale density map of the distribution of effective interaction fields experienced by a single cell $h_{\rm i}^{\rm int}$ versus its own bias or local field $h_{\rm i}$ (see text); distribution formed over network configurations at each point in time during a natural movie for n=1140 3--cell groups (top panel) and n=250 10--cell groups (bottom panel).  Black line shows the boundary between dominance of local fields vs. interactions.
(b) Mean interactions $J_{\rm ij}$ and local fields $h_{\rm i}$ describing groups of  $N$ cells,  with error bars showing standard deviations across multiple groups. While local fields decline significantly with increasing network size, interaction strengths do not.  (c) Pairwise interaction in a network of 10 cells $J_{\rm ij}^{10}$ plotted against the interaction values of the same pair in a sub-network containing only 5 cells $J_{\rm ij}^5$. Line shows equality.}

\end{centering}
\end{figure}

\section{Larger networks and error correction}

Groups of $N=10$ cells  are large enough to reveal dramatic departures from independence, but  small enough that we can directly sample the relevant probability distributions.  What happens at larger $N$? 
In general we expect that the total capacity of the network to represent its sensory inputs should grow in proportion to the number of neurons $N$.      This is the usual thermodynamic limit in statistical mechanics, where energy and entropy are proportional to system size \cite{LL}.  But this behavior is not guaranteed when all elements of the system interact with each other.  In the Ising model, it is known that if all pairs of spins (here, cells) interact significantly, then to recover the thermodynamic limit
the typical size of the interactions $J_{\rm ij}$ must decrease with $N$ \cite{amit_89,frust}. Although we have not analyzed very large networks, we see no signs of significant changes in $J$ with growing $N$  (Fig 4b, c).

In a physical system,  the maximum entropy distribution is equivalent to the Boltzmann distribution  $P\propto \exp(-E/k_B T)$, and  the behavior of the system  depends on the temperature $T$.  For the network of neurons, there is no real temperature, but the statistical mechanics of the Ising model predicts that when all pairs of elements interact, increasing the number of elements while fixing the typical strength of interactions is
equivalent to lowering the temperature $T$ in a physical system of fixed size $N$.
This mapping predicts that correlations will be even more important in larger groups of neurons.

We can see signs of strong correlation emerging by looking at the entropy and multi--information in networks of different sizes.  If all cells were independent, the entropy would be $S_1$,  exactly  proportional to $N$.   For weak correlations, we can solve the Ising model in perturbation theory to show that the multi--information $I_N$ is the sum of mutual information terms between all pairs of cells and hence $I_N \propto N^2$. This is in agreement with the  empirically estimated $I_N$ up to $N=15$, the largest value for which direct sampling of the data provides a good estimate (Fig 5a), and Monte Carlo simulations of the maximum entropy models suggest that this agreement extends up to the full population of $N=40$ neurons in our experiment (G Tka\v{c}ik, ES, RS, MJB \& WB, unpublished).  Were this pattern to continue, at $N\sim 200$ cells $I_N$ would become equal to the independent entropy $S_1$, and the true entropy $S_N = S_1 - I_N$ would vanish as the system ``froze.''    Because we see variable firing patterns of all the cells, we know that literal freezing of the network into a single state doesn't happen. On the other hand,  networks of $N\sim 200$ cells must be very strongly ordered.
Interestingly, experiments indicate that a correlated patch on the retina has roughly this size: the strongest correlations are found for cells within $\sim 200\,\mu{\rm m}$ of each other, and this area contains $\sim 175$ ganglion cells in the salamander \cite{data_b}.

Because the interactions $J_{\rm ij}$ have different signs, frustration can prevent the freezing of the system into a single state.  Instead there will be multiple states that are local minima of the effective energy function, 
as in spin glasses  \cite{frust}.
If the number of minimum energy patterns is not too small, then the system retains a significant representational capacity.  If the number of patterns is not too large, then observing only some of the cells in the network is sufficient to identify the whole pattern uniquely, just as in the Hopfield model of associative memory \cite{hopfield_82}.  Thus the system would have a holographic or error--correcting property, so that an observer who has access only to a fraction of the neurons would nonetheless be able to reconstruct the activity of the whole population. 

\begin{figure}
\vskip -1.1 in
\centerline{\psfig{figure=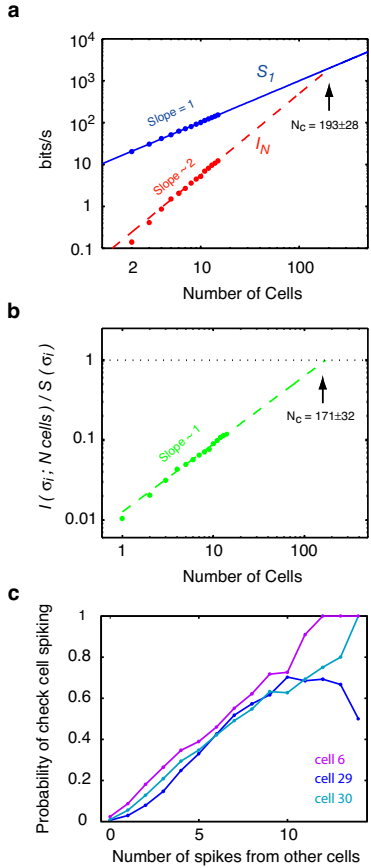,width=1.6\columnwidth}}
\vskip -1.1 in
\caption{Extrapolation to larger networks. (a) Average independent cell entropy $S_1$ and  network multi-information $I_N$, multipled by $1/\Delta \tau$ to give bin--independent rates, vs. number of cells in the network $N$. Theoretically we expect $I_N \propto N(N-1)$ for small $N$; the best fit is $I_N \propto N^{1.98\pm 0.04}$.  Extrapolating  (dashed line) defines a critical network size $N_c$, where $I_N$ would be equal to $S_1$. (b)  Information that $N$ cells provide about the activity of the $N+1^{\rm st}$ cell, plotted as a fraction of that cell's entropy, vs. network size $N$; each point is the average value for many different groups of cells. Extrapolation to larger networks (dashed line, slope = $1.017\pm 0.052$) defines another critical network size $N_c$, where one would get perfect error--correction or prediction of the state of a single cell from the activity of the rest of the network. (c) Examples of `check cells,' for which the probability of spiking is an almost perfectly linear encoding of the number of spikes generated by the other cells in the network. Cell numbers as in Fig 1.}
\end{figure}

We can see suggestions of this error--correcting property by asking
directly how much the knowledge of activity in $N$ cells tells us about the whether the $N+1^{\rm st}$ cell will spike (Fig 5b). Our uncertainty about the state of one cell is reduced in proportion to the number of cells that we examine, and  if this trend continues then again at $N \sim 200$ all uncertainty would vanish.   Alternatively, we can look for particular kinds of error correction.  In our population of 40 cells we have found three cells for which the probability of spiking is an almost perfectly linear encoding of the number of spikes generated by the other cells in the network (Fig 5c); by observing the activity of these ``check cells'' we can estimate how many spikes are generated by the network as a whole even before we observe any of the other cells' responses.

\section{Discussion}

Returning to the general questions raised at the outset, we have seen that the maximum entropy principle provides a unique candidate model for the whole network that is consistent with observations on pairs of elements but makes no additional assumptions.  Despite the opportunity for higher--order interactions in the retina, this model captures more than 90\% of the structure in the detailed patterns of spikes and silence in the network, closing the enormous gap between the data and the predictions of a model in which cells fire independently.  Because the maximum entropy model 
has relatively few parameters, we evade the curse of dimensionality and associated sampling problems that would ordinarily limit the exploration of larger networks.

Key ingredients in matching theory to experiment are the focus on time windows in which the discreteness of the neural responses is evident,  low spike probabilities, and  weak but significant correlations among almost all pairs of cells. 
None of these considerations are specific to the retina.
As a first test for the generality of thee ideas,   we have analyzed experiments on cultured networks of cortical neurons \cite{marom}, where again we find that the maximum entropy model captures over 95\% of the multi--information in groups of $N=10$ cells (Fig. 2d).

The success of a model that includes only pairwise interactions 
provides an enormous simplification in our description of the network.  This may be important not just for us, but for the brain as well.
The dominance of pairwise interactions means that learning rules based on pairwise correlations \cite{hebb} could be sufficient  to generate nearly optimal internal models for the distribution of ``codewords'' in the retinal vocabulary,
thus allowing the brain  to accurately evaluate new events for their degree of surprise \cite{barlow}.


The mapping of the maximum entropy problem to the Ising model, together with the observed level of correlations, implies that 
groups of $N\sim 200$ cells will behave very differently than smaller groups,
and this is especially interesting because  the patch of significantly correlated ganglion cells in the retina is close to this critical size \cite{data_b}.
Because the response properties of retinal ganglion cells adapt to  the input image statistics \cite{smirnakis+al_97,hosoya+al_05}, this matching of correlation length and correlation strength cannot be an accident of anatomy but rather must be set by adaptive mechanisms.  Perhaps there is an optimization principle which determines this operating point, maximizing coding capacity while maintaining the correlation structures which enable error--correction.

Although we have focused on networks of neurons, the same framework has the potential to describe biological networks more generally. In this view, the network is much more than the sum of its parts, but a nearly complete model can be derived from all of its pairs.

\section*{Methods}

{\bf Electrophysiological Recording.}  Retinae from the larval tiger salamander ({\em Ambystoma tigrinum}) and the guinea pig ({\em Cavia porcellus}) were isolated from the eye retaining the pigment epithelium and placed over a multi-electrode array \cite{data_b}.  Both were perfused with oxygenated medium: room temperature RingerÕs for salamander and $36^\circ$C Ames medium for guinea pig.  Extracellular voltages were recorded by a MultiChannel Systems MEA 60 microelectrode array and streamed to disk for offline analysis.  Spike waveforms were sorted either using the spike size and shape from a single electrode \cite{data_b} or the full waveform on 30 electrodes \cite{data_a}.  Recorded ganglion cells were spaced no more than $500\,\mu{\rm m}$  apart, and were typically close enough together to have overlapping receptive field centers.  This analysis presented here is based on measurements of 95 cells recorded in 4 salamanders and 35 cells recorded in 2 Guinea pigs.

{\bf Visual Stimulation.} Natural movie clips (NAT in Fig. 3) were acquired using a Canon Optura Pi video camera at 30 frames per second.  Movie clips broadly sampled woodland scenes as well as man--made environments, and included several qualitatively different kinds of motion: objects moving in a scene, optic flow, and simulated saccades; see Ref. 19 
for  details.  In spatially uniform flicker (FFF), the light intensity was randomly chosen to be black or white every 16.7 ms.  For most experiments, a 20--30 s stimulus segment was repeated many times; in one experiment, a 16 min movie clip was repeated several times.  All visual stimuli were displayed on an NEC FP1370 monitor and projected onto the retina using standard optics.  The mean light level was 5 Lux, corresponding to photopic vision.    

{\bf Cultured cortical networks.} 
Data on cultured cortical neurons were recorded by
the laboratory of S. Marom (Technion--Israel Institute of Technology)
using a multi--electrode array, as described in  Ref. \cite{marom}. 
The data set analyzed here is an hour long epoch of
spontaneous neuronal activity recorded through 60 electrodes.

{\bf Estimating information theoretic quantities.}   Information theoretic quantities such as  $I_N$ depend on the full distribution of states for the real system.  Estimating these quantities can be difficult, because finite data sets lead to systematic errors  \cite{workshop}.  With large data sets ($\sim 1\,{\rm hr}$) and $N<15$ cells, however, systematic errors are small, and we can use the sample--size dependence of the estimates to correct for these errors, as in Ref. \cite{strong+al_98}.

{\bf Constructing the maximum entropy distributions.}
For networks of modest size, as considered here, constructing the maximum entropy distribution consistent with the mean spike rates and pairwise correlations can be viewed as an optimization problem with constraints.  Because the entropy is a convex function of the probabilities, and the constraints are linear, 
many efficient algorithms are available \cite{GIS}.

\acknowledgments {We thank G. Stephens and G. Tkacik for discussions, N.  Tkachuk for help with the experiments,  and S. Marom for sharing his lab's cultured cortical networks data with us.  This work was supported in part by NIH  Grants R01 EY14196 and P50 GM071508, and by the E. Matilda Zeigler Foundation.
}

\end{document}